\documentclass[aps,pra,12pt,onecolumn,superscriptaddress]{revtex4-1}

\usepackage{amsmath,amsfonts,amssymb,color,epsfig,graphics,graphicx,latexsym,revsymb,theorem,url,verbatim,epstopdf}

\usepackage{hyperref}

\newtheorem{definition}{Definition}
\newtheorem{proposition}[definition]{Proposition}
\newtheorem{lemma}[definition]{Lemma}

\newtheorem{theorem}[definition]{Theorem}
\newtheorem{corollary}[definition]{Corollary}
\newtheorem{conjecture}[definition]{Conjecture}

\newtheorem{remark}[definition]{Remark}
\newtheorem{example}[definition]{Example}
\newtheorem{question}[definition]{Question}

\def\squareforqed{\hbox{\rlap{$\sqcap$}$\sqcup$}}
\def\qed{\ifmmode\squareforqed\else{\unskip\nobreak\hfil
\penalty50\hskip1em\null\nobreak\hfil\squareforqed
\parfillskip=0pt\finalhyphendemerits=0\endgraf}\fi}
\def\endenv{\ifmmode\;\else{\unskip\nobreak\hfil
\penalty50\hskip1em\null\nobreak\hfil\;
\parfillskip=0pt\finalhyphendemerits=0\endgraf}\fi}
\newenvironment{proof}{\noindent \textbf{{Proof.~} }}{\qed}
\def\Dbar{\leavevmode\lower.6ex\hbox to 0pt
{\hskip-.23ex\accent"16\hss}D}
\makeatletter
\def\url@leostyle{%
  \@ifundefined{selectfont}{\def\UrlFont{\sf}}{\def\UrlFont{\small\ttfamily}}}
\makeatother
\def\bcj{\begin{conjecture}}
\def\ecj{\end{conjecture}}
\def\bcr{\begin{corollary}}
\def\ecr{\end{corollary}}
\def\bd{\begin{definition}}
\def\ed{\end{definition}}
\def\bea{\begin{eqnarray}}
\def\eea{\end{eqnarray}}
\def\bem{\begin{enumerate}}
\def\eem{\end{enumerate}}
\def\bex{\begin{example}}
\def\eex{\end{example}}
\def\bim{\begin{itemize}}
\def\eim{\end{itemize}}
\def\bl{\begin{lemma}}
\def\el{\end{lemma}}
\def\bpf{\begin{proof}}
\def\epf{\end{proof}}
\def\bpp{\begin{proposition}}
\def\epp{\end{proposition}}
\def\bqu{\begin{question}}
\def\equ{\end{question}}
\def\br{\begin{remark}}
\def\er{\end{remark}}
\def\bt{\begin{theorem}}
\def\et{\end{theorem}}

\def\btb{\begin{tabular}}
\def\etb{\end{tabular}}

\newcommand{\nc}{\newcommand}

\def\a{\alpha}
\def\b{\beta}

\def\z{\zeta}

\def\x{\xi}

\def\r{\rho}
\def\s{\sigma}

\def\ph{\varphi}

\def\ps{\psi}

\def\G{\Gamma}

\def\T{\Theta}

 \nc{\bbA}{\mathbb{A}} \nc{\bbB}{\mathbb{B}} \nc{\bbC}{\mathbb{C}}
 \nc{\bbD}{\mathbb{D}} \nc{\bbE}{\mathbb{E}} \nc{\bbF}{\mathbb{F}}
 \nc{\bbG}{\mathbb{G}} \nc{\bbH}{\mathbb{H}} \nc{\bbI}{\mathbb{I}}
 \nc{\bbJ}{\mathbb{J}} \nc{\bbK}{\mathbb{K}} \nc{\bbL}{\mathbb{L}}
 \nc{\bbM}{\mathbb{M}} \nc{\bbN}{\mathbb{N}} \nc{\bbO}{\mathbb{O}}
 \nc{\bbP}{\mathbb{P}} \nc{\bbQ}{\mathbb{Q}} \nc{\bbR}{\mathbb{R}}
 \nc{\bbS}{\mathbb{S}} \nc{\bbT}{\mathbb{T}} \nc{\bbU}{\mathbb{U}}
 \nc{\bbV}{\mathbb{V}} \nc{\bbW}{\mathbb{W}} \nc{\bbX}{\mathbb{X}}
 \nc{\bbZ}{\mathbb{Z}}

 \nc{\bA}{{\bf A}} \nc{\bB}{{\bf B}} \nc{\bC}{{\bf C}}
 \nc{\bD}{{\bf D}} \nc{\bE}{{\bf E}} \nc{\bF}{{\bf F}}
 \nc{\bG}{{\bf G}} \nc{\bH}{{\bf H}} \nc{\bI}{{\bf I}}
 \nc{\bJ}{{\bf J}} \nc{\bK}{{\bf K}} \nc{\bL}{{\bf L}}
 \nc{\bM}{{\bf M}} \nc{\bN}{{\bf N}} \nc{\bO}{{\bf O}}
 \nc{\bP}{{\bf P}} \nc{\bQ}{{\bf Q}} \nc{\bR}{{\bf R}}
 \nc{\bS}{{\bf S}} \nc{\bT}{{\bf T}} \nc{\bU}{{\bf U}}
 \nc{\bV}{{\bf V}} \nc{\bW}{{\bf W}} \nc{\bX}{{\bf X}}
 \nc{\bZ}{{\bf Z}}

\nc{\cA}{{\cal A}} \nc{\cB}{{\cal B}} \nc{\cC}{{\cal C}}
\nc{\cD}{{\cal D}} \nc{\cE}{{\cal E}} \nc{\cF}{{\cal F}}
\nc{\cG}{{\cal G}} \nc{\cH}{{\cal H}} \nc{\cI}{{\cal I}}
\nc{\cJ}{{\cal J}} \nc{\cK}{{\cal K}} \nc{\cL}{{\cal L}}
\nc{\cM}{{\cal M}} \nc{\cN}{{\cal N}} \nc{\cO}{{\cal O}}
\nc{\cP}{{\cal P}} \nc{\cQ}{{\cal Q}} \nc{\cR}{{\cal R}}
\nc{\cS}{{\cal S}} \nc{\cT}{{\cal T}} \nc{\cU}{{\cal U}}
\nc{\cV}{{\cal V}} \nc{\cW}{{\cal W}} \nc{\cX}{{\cal X}}
\nc{\cZ}{{\cal Z}}

\nc{\hA}{{\hat{A}}} \nc{\hB}{{\hat{B}}} \nc{\hC}{{\hat{C}}}
\nc{\hD}{{\hat{D}}} \nc{\hE}{{\hat{E}}} \nc{\hF}{{\hat{F}}}
\nc{\hG}{{\hat{G}}} \nc{\hH}{{\hat{H}}} \nc{\hI}{{\hat{I}}}
\nc{\hJ}{{\hat{J}}} \nc{\hK}{{\hat{K}}} \nc{\hL}{{\hat{L}}}
\nc{\hM}{{\hat{M}}} \nc{\hN}{{\hat{N}}} \nc{\hO}{{\hat{O}}}
\nc{\hP}{{\hat{P}}} \nc{\hR}{{\hat{R}}} \nc{\hS}{{\hat{S}}}
\nc{\hT}{{\hat{T}}} \nc{\hU}{{\hat{U}}} \nc{\hV}{{\hat{V}}}
\nc{\hW}{{\hat{W}}} \nc{\hX}{{\hat{X}}} \nc{\hZ}{{\hat{Z}}}

\nc{\hn}{{\hat{n}}}

\def\dim{\mathop{\rm Dim}}
\def\End{\mathop{\rm End}}

\def\max{\mathop{\rm max}}

\def\rank{\mathop{\rm rank\;}}

\def\tr{\mathop{\rm Tr}}

\def\GL{{\mbox{\rm GL}}}

\def\ox{\otimes}

\def\pars{\partial\cS}

\def\su{\subset}
\def\sue{\subseteq}

\newcommand{\bra}[1]{\langle#1|}
\newcommand{\ket}[1]{|#1\rangle}
\newcommand{\proj}[1]{| #1\rangle\!\langle #1 |}
\newcommand{\ketbra}[2]{|#1\rangle\!\langle#2|}

\newcommand{\opp}{\red{OPEN PROBLEMS}.~}

\newcommand{\red}{\textcolor{red}}

\newcommand{\cmp}{Commun. Math. Phys.}

\newcommand{\jmp}{J. Math. Phys.}

\begin{document}
\title{Length filtration of the separable states}

\author{Lin Chen}
\email{linchen@buaa.edu.cn (corresponding author)}
\affiliation{School of Mathematics and Systems Science, Beihang University, Beijing 100191, China}
\affiliation{International Research Institute for Multidisciplinary Science, Beihang University, Beijing 100191, China}

\def\Dbar{\leavevmode\lower.6ex\hbox to 0pt
{\hskip-.23ex\accent"16\hss}D}
\author {{ Dragomir {\v{Z} \Dbar}okovi{\'c}}}
\email{djokovic@uwaterloo.ca}
\affiliation{Department of Pure Mathematics and Institute for
Quantum Computing, University of Waterloo, Waterloo, Ontario, N2L
3G1, Canada}

\begin{abstract}
We investigate the separable states $\r$ of an arbitrary 
multipartite quantum system with Hilbert space $\cH$ of dimensionin $d$. The length $L(\r)$ of $\r$ is defined 
as the smallest number of pure product states having $\r$ as 
their mixture. The length filtration of the set of separable states, $\cS$, is the increasing chain 
$\emptyset\subset\cS'_1\subseteq\cS'_2\subseteq\cdots$, where
$\cS'_i=\{\r\in\cS:L(\r)\le i\}$.
We define the maximum length, 
$L_{\rm max}=\max_{\r\in\cS} L(\r)$, critical length, 
$L_{\rm crit}$, and yet another special length, $L_c$, which 
was defined by a simple formula in one of our previous papers. 
The critical length indicates the first term in the length 
filtrartion whose dimension is equal to $\dim\cS$. 
We show that in general
$d\le L_c\le L_{\rm crit}\le L_{\rm max}\le d^2$.

We conjecture that 
the equality $L_{\rm crit}=L_c$ holds for all finite-dimensional 
multipartite quantum systems. 
Our main result is that $L_{\rm crit}=L_c$ for the bipartite 
systems having a single qubit as one of the parties. This is 
accomplished by computing the rank of the Jacobian matrix of 
a suitable map having $\cS$ as its range. 
\end{abstract}

\date{\today}


\maketitle

\tableofcontents

\section{Introduction}\label{sec1}

In quantum information theory, quantum entanglement is the basic resource and quantum separable states do not contain entanglement \cite{werner89}. Understanding the properties of separable states and deciding whether a given state is separable (an NP-hard problem) is one of the fundamental open problems of quantum physics. By the entanglement measure for mixed quantum states \cite{dtt00}, the length of separable states represents the minimal physical effort needed to implement the state. 
Two separable states of different lengths are not equivalent under stochastic local operations and classical communications \cite{dvc00}. Further, the length of the operator of the bipartite symmetric subspace is related to the existence of symmetrically-informational-completely positive operator-valued measure (SIC-POVM) \cite{scott06}, which is a main open problem in quantum measurement and information. In spite of the various applications of length, its computation is mathematically hard and has attracted much attention in recent years \cite{uhlmann97,as10,cd12pra,hk13,cd13jmp,cd13jpa,cd13,cd14,hk14,hk14cmp}.

To state and explain our results we need the following definitions which, will be used in the whole paper. Let 
$\cH=\cH_1\otimes\cH_2\otimes\cdots\otimes\cH_n$ be the
complex Hilbert space of a finite-dimensional $n$-partite quantum
system. We denote by $d_i$ the dimension of $\cH_i$, and so
$d:=\prod d_i$ is the dimension of $\cH$. To avoid trivial cases, we assume that each $d_i>1$ and $n>1$. 
A {\em product vector} is a nonzero vector of the form $\ket{x}=\ket{x_1}\ox\cdots\ox\ket{x_n}$ where $\ket{x_i}\in\cH_i$. We shall write this product vector also as $\ket{x_1,\ldots,x_n}$. 
A {\em pure product state} is a state $\r$ of the form $\r=\proj{x}$ where $\ket{x}$ is a product vector. 

A (non-normalized) state $\r$ is {\em separable} if it is a sum of pure product states, i.e., 
\begin{equation} \label{eq:covex-dec}
\r=\sum_{k=1}^l \proj{z_k}, 
\end{equation}
where the $\ket{z_k}$ are product vectors. The {\em length}, 
$L(\r)$, of $\r$ is the smallest integer $l$ over all such decompositions of $\r$. 

In this paper, we will investigate the separable states in terms of their rank and length, see \eqref{eq:S_r}. They provide two filtrations of the set of separable states, namely the rank and length filtration, see \eqref{R-filtration} and 
\eqref{L-filtration}. 
Some inclusion relations among the first few terms of these two 
filtrations are proved in \eqref{eq:cs'nsub} and Proposition \ref{pp:kleq4}. To further investigate the length filtration we introduce the notions of maximum length, critical length and recall an older specific length $L_c$ in Definition \ref{df:ml}. Their relation is elucidated in Proposition \ref{pp:lclcrit} and Conjecture \ref{cj:Prva-hip}. This conjecture is the main problem of this paper, and we will prove that it is true for the 
bipartite $2\ox d_2$ systems, see Theorem \ref{thm:glavna}. 
For this purpose, we define the map $\Phi_r$ in \eqref{eq:Phi_r}, and compute the rank of its Jacobian matrix 
for $r=d$, see Sec. \ref{sec:Jacobian}.

A vector $\ket{x}\in\cH$ is {\em normalized} if $\|x\|=1$. We denote by $H$ the space of Hermitian operators $\r$ on $\cH$. Note that $H$ is a real vector space of dimension $d^2$. We denote by $H_1$ the affine hyperplane of $H$ defined by the equation $\tr\r=1$.
The mixed quantum states of this quantum system are represented
by their density matrices, i.e., operators $\r\in H$ which are
positive semidefinite $(\r\ge0)$ and have unit trace
$(\tr \r=1)$. For convenience, we often work with non-normalized
states, i.e., Hermitian operators $\r$ such that $\r\ge0$ and
$\r\ne0$. It will be clear from the context whether we require
the states to be normalized. We denote by $\cR(\r)$ the range of a linear operator $\r$.

We assume that an orthonormal basis is fixed in each $\cH_i$
and we use the standard notation $\ket{0},\ldots,\ket{d_i-1}$
for the corresponding basis vectors. The product vectors 
$\ket{i_1,i_2,\ldots,i_n}$, $0\le i_k<d_k$, form an orthonormal (o.n.) basis of $\cH$. We refer to this basis as the {\em standard basis}. When necessary, we 
shall write the standard basis vector $\ket{i}\in\cH_q$ as
$\ket{i}_q$. We write $\End V$ for the algebra of linear operators on a finite-dimensional complex vector space $V$. 
The operation of transposition applied only to the $i$th tensor factor of
$\End\cH=\End\cH_1\ox\cdots\ox\End\cH_i\ox\cdots\ox\End\cH_n$ 
will be denoted by $\G_i$. We denote by $\Theta$ the abelian group of order $2^n$ generated by the $\G_i$s. We refer to the elements of $\T$ as the {\em partial transposition operators}. Thus if $\r$ is a state on $\cH$, then $\G_i\r$ is the $i$th partial transpose of $\r$.

A vector $\ket{x}\in\cH$ is {\em real} if all components of 
$\ket{x}$ (with respect to the standard basis) are real. 
A state is {\em real} if its density matrix is a real 
symmetric matrix.
The group $\Theta$ preserves the length of separable states 
$\r$, i.e., we have 
\bea \label{eq:LG}
L(\G\r)=L(\r),\quad \forall \G\in\Theta.
\eea
On the other hand $\r$ and $\G\r$ may have different ranks.
In the bipartite case, for any linear operator $\r$ on $\cH$, 
we refer to the ordered pair 
$(\rank\r,\rank\G_1\r)$ as the {\em birank} of $\r$.


\section{Two filtrations of the set of separable states}

We denote by $\cS$ the set of normalized separable states. 
Let $\pars$ denote the boundary of $\cS$. 
For any subset $X\sue\cS$ let $KX$ be the cone over $X$, i.e., $KX=\{t\r:t\ge0,~\r\in X\}$.
For any positive integer $r$ we set 
\bea \label{eq:S_r}
\cS_r=\{\r\in\cS:\rank\r\le r\} \quad {\rm and} \quad 
\cS'_r=\{\r\in\cS:L(\r)\le r\}.
\eea

Obviously $\cS'_i\sue\cS'_{i+1}$ for each $i$, 
and it is well known that $\cS'_{d^2}=\cS$. 
It is also known that $\dim\cS=d^2-1$ \cite[Theorem 1]{ZHSL98}.
Note that $\cS_{d-1}$ is contained in the hypersurface of 
$H_1$ defined by the equation $\det\r=0$. Consequently, we have
\bea \label{eq:dimS_d-1}
\dim \cS_{d-1} \le d^2-2.
\eea
 
Since for each $k\le d$ there exists $\r\in\cS'_k$ having rank 
$k$, we have
\bea \label{eq:S'_d}
\cS'_k \not\sue \cS_{k-1} \quad \text{for} \quad k\le d.
\eea

The dimensions of the sets $\cS'_k$ for all $k$ were computed 
for several systems in our paper \cite{cd13jmp}. In particular  
this was done for bipartite systems $2\ox N$ with $N<9$. 
We have extended these computations to all $N\le20$. Hence the 
results presented in \cite[Table I]{cd13jmp} for the $2\ox N$ 
case are valid in the extended range $1<N\le20$. In particular, 
in these cases we have 
$\dim\cS'_d=d^2-1$ and $\dim\cS'_{d-1}=d^2-3$,

It is much harder to compute the dimension of the sets $\cS_k$.
For instance, in the two-qubit case we know that 
$\dim \cS_1=4$, $\dim \cS_2=9$ and $\dim \cS_4=15$ because 
$\cS_1=\cS'_1$, $\cS_2=\cS'_2$ and $\cS_4=\cS$. 
It follows from \eqref{eq:dimS_d-1} that $\dim \cS_3\le 14$ 
and since $\dim \cS'_3=13$ (see \cite[Table I]{cd13jmp}) 
we have $\dim \cS_3\ge 13$. The following general lemma implies 
that this dimension is 14. 

\bl \label{le:raste}
If $\dim\cS'_k<d-1$ then $\dim\cS'_k<\dim\cS_{k+1}$.
\el
\bpf
Since $\cS'_k$ is a semialgebraic set, it is a finite disjoint 
union of $C^\infty$ submanifolds of $H_1$. At least one of these 
submanifolds, say $X$, has dimension equal to $m:=\dim\cS'_k$. 
Fix a point $\s\in X$ and choose a pure product state $\r$ 
not in the tangent plane to $X$ at $\s$. Then the union $Y$ of 
all line segments joining $\r$ to a point of $X$ has dimension 
$m+1$. As $Y\sue\cS_{k+1}$, we have 
$\dim\cS_{k+1}\ge\dim Y>m$.
\epf

More generally, this lemma implies that the equality sign holds 
in \eqref{eq:dimS_d-1} in the case of $2\ox N$ systems with 
$1<N\le20$.

The sets defined in \eqref{eq:S_r} form two filtrations of 
$\cS$:
\bea
\label{R-filtration}
&& \emptyset\su\cS_1\su\cS_2\su\cdots\su\cS_d=\cS, \\
\label{L-filtration}
&& \emptyset\su\cS'_1\sue\cS'_2\sue\cdots\sue\cS'_{d^2}=\cS.
\eea

We refer to them as the {\em rank filtration} and the 
{\em length filtration} of $\cS$, respectively.

It is easy to see that $\cS_k$ and $\cS'_k$ are closed sets.
We remark that in fact $\cS_k$ is the closure of the set 
$\{\r\in\cS:\rank(\r)=k\}$, and $\cS'_k$ is the closure of 
$\{\r\in\cS:L(\r)=k\}$.

Let us compare the initial terms of these two filtrations.
Note that $\cS'_1=\cS_1$ and that $\cS'_k\sue\cS_k$ for each 
$k$. 
It follows from \cite[Lemma 11]{cd13jpa} that also 
$\cS'_2=\cS_2$. On the other hand, we claim that 
\bea \label{eq:S_3}
\cS'_3\su\cS_3\su\cS'_4. 
\eea
Since there exist separable states 
of rank 3 and length 4, see \cite[Table I]{cd12pra}, 
we have $\cS'_3\su\cS_3$. The fact that $\cS_2=\cS'_2$ and 
\cite[Theorem 15]{cd13jpa} imply that $\cS_3\sue\cS'_4$. 
Since there exist separable states of rank 4 and length 4, this 
inclusion is strict and our claim is proved.

Next we claim that for $d>4$ we have
\bea \label{eq:S_4}
\cS'_4\sue\cS_4\su\cS'_6. 
\eea
The first inclusion is obvious. 
It follows from \cite[Lemma 17]{cd13jpa} that $\cS_4\sue\cS'_6$. So the second inclusion relation is equivalent to show that there is some state in $\cS'_6$ but not in $\cS_4$. 
The state can be chosen as the partial transpose of the state in 
\cite[Eq. (14)]{cd13jpa}, a 3-qubit state of rank six 
and length six.

For $\r\in\cS'_k$ and $\G\in\Theta$ we have 
$\rank \G\r\le L(\G\r)=L(\r)\le k$. Hence,
\bea 
\label{eq:cs'nsub}
\cS'_k \subseteq \{ \r\in\cS_k:
\rank\Gamma\r\le k,~\forall\Gamma\in\Theta\}.
\eea

If equality sign holds in \eqref{eq:cs'nsub}, then we obtain a 
very simple characterization of $\cS'_k$ as a subset of $\cS_k$.
We now that the equality holds if and only if $k\le4$.

\bpp
\label{pp:kleq4}
Let $\r$ be a multipartite separable state. Then

(i) $L(\r)=3$ if and only if $\rank\G\r=3$, $\forall \G\in\T$.

(ii) The equality sign holds in \eqref{eq:cs'nsub} if and 
only if $k\le 4$.
\epp
\bpf
(i) If $L(\r)=3$ then \eqref{eq:cs'nsub} shows that 
$\rank\G\r\le3$, $\forall \G$. If $\rank\G\r<3$ for some $\G$ then  $\G\r\in\cS_2=\cS'_2$, contradicting that $L(\G\r)=3$. 
Conversely, assume that $\rank\G\r=3$, $\forall\G$. 
Then \cite[Theorem 15]{cd13jpa} shows that $L(\r)$ is 3 or 4,
and if it is 4 then $\r$ is a two-qubit state. The possibility 
$L(\r)=4$ is ruled out by \cite[Table 1]{cd12pra}.

(ii) Let $\r\in\cS_k$ be such that $\rank\G\r\le k$, 
$\forall\G\in\T$. We have to prove that $\r\in\cS'_k$ if 
$k\le4$. If $k=1$ or 2 then $\r\in\cS_k=\cS'_k$.

Now let $k=3$. If $\rank\G\r<3$ for some $\G$, then 
$\G\r\in\cS_2=\cS'_2$ and so $\r\in\cS'_2\su\cS'_3$. 
Thus we may assume that $\rank\G\r=3$, $\forall \G$. 
Then (i) shows that $\r\in\cS'_3$.

Finally, let $k=4$. Assume that $\rank\G\r<4$ for some $\G$, 
i.e., that $\G\r\in\cS_3$.
If $\G\r\in\cS_2=\cS'_2$ then $\r\in\cS'_2\sue\cS'_4$. 
Otherwise $\rank\G\r=3$ and \cite[Theorem 15]{cd13jpa} shows 
that $\G\r\in\cS'_4$. Consequently $\r\in\cS'_4$.
From now on we assume that $\rank\G\r=4$, $\forall\G$. 

Assume that $\r$ is $A_i$-reducible for some index $i$, i.e., 
$\r=\a\oplus_{A_i}\b$, see \cite[Definition 6]{cd13jpa} 
for the definition of reducibility and irreducibility. 
It follows that $\G\r=\G\a\oplus_{A_i}\G\b$, $\forall\G$. 
Consequently, both $\G\a$ and $\G\b$ have rank at most 3. 
Since (iii) is already proved for $k\le3$, we conclude that 
$L(\a)=\rank\a$ and $L(\b)=\rank\b$. It follows that 
$L(\r)=4$, and so $\r\in\cS'_4$.
From now on we assume that $\r$ is irreducible.

Let $r_i$ denote the rank of the reduced density operator 
$\r_{A_i}$. We may assume that $r_1\le r_2\le\cdots\le r_n$.

Let us first consider the bipartite case $(n=2)$. Since $\r$ 
is irreducible, we have $r_1>1$. Thus $2\le r_1\le r_2\le4$. 
If $r_2=2$ then $\r\in\cS'_4$ because all separable 
two-qubit states have length at most 4. 
If $r_2=3$ then $\r\in\cS'_4$ by \cite[Proposition 3]{cd12pra}. 
If $r_2=4$ then $\r\in\cS'_4$ by \cite[Corollary 3(a)]{kck00}.  

Now let $n>2$. Since $\r$ is irreducible, 
\cite[Lemma 17 (ii)]{cd13jpa} implies that $L(\r)\le4$ when $r_n>2$. Hence the assertion holds. We have 
$r_i=2$ for all $i$.
In the paragraphs from the beginning to $(13)$ in the proof of \cite[Lemma 17 (iii)]{cd13jpa}, we have shown that $L(\r)\le4$, except that $\r=\sum^l_{i=1} \proj{a_i,\cdots,a_i}$ with $l\ge5$. Furthermore, we have the fact that any $k$-partite reduced density operator of $\r$ has rank three for $k\in[2,n-1]$. Up to ILOs we can assume that 
$\ket{a_1}=\ket{0}$ and $\ket{a_2}=\ket{1}$.
By replacing $\r$ by $\G\r$ with any $\G\in\T$ in the above argument, we can assume that the $\ket{a_i}$ are all real and pairwise linearly independent. 
If $n>3$, the tripartite reduced density operator of $\r$ has rank bigger than three. It gives us a contradiction with the above fact. 

So $n=3$, i.e.,
$\r=\proj{0,0,0}+\proj{1,1,1}+\sum^l_{i=3} \proj{a_i,a_i,a_i}$ where $l\ge5$, the $\ket{a_i}$ are all real and pairwise linearly independent. We regard $\r$ as a bipartite state with the system split $A_1:A_2A_3$. So $\r$ is a $2\times3$ separable state of birank $(4,4)$. It follows from \cite{cd12pra} that $\r=\sum^4_{j=1}\proj{b_j,c_j}$ where $\ket{c_j}$ is a two-qubit state of the system $A_2A_3$. Any $\ket{b_j,c_j}$ is in the range of $\r$, which is the 3-qubit symmetric subspace, So $\ket{c_j}\propto\ket{b_j,b_j}$, and $L(\r)\le4$. 

Thus we have proved the ``if'' part of (ii). The ``only if'' part 
follows from the example below. This completes the proof.
\epf

\bex \label{ex:Tiles}
{\rm
We construct a $3\times3$ separable state $\r$ of length 6 
such that both $\r$ and $\Gamma_1\r$ have rank 5. 
The product vectors
\bea
 \label{eq:3x3upb}
 \ket{\ps_1}&=&\frac{1}{\sqrt2}\ket{0}(\ket{0}-\ket{1}),
 \nonumber\\
 \ket{\ps_2}&=&\frac{1}{\sqrt2}\ket{2}(\ket{1}-\ket{2}),
 \nonumber\\
 \ket{\ps_3}&=&\frac{1}{\sqrt2}(\ket{0}-\ket{1})\ket{2},
 \nonumber\\
 \ket{\ps_4}&=&\frac{1}{\sqrt2}(\ket{1}-\ket{2})\ket{0},
 \nonumber\\
 \ket{\ps_5}&=&\frac13(\ket{0}+\ket{1}+\ket{2})(\ket{0}+\ket{1}+\ket{2})
\eea
form an unextendible product basis (UPB) \cite{bdm99}.
There is a unique sixth product vector in their span, namely
\bea
 \label{ea:3x3upb,constant6th}
 \ket{\ps_6}
 &=&
 \frac19 (2\ket{0}-\ket{1}+2\ket{2})(2\ket{0}-\ket{1}+2\ket{2})
 \nonumber\\
 &=&
 {1\over3}(\ket{\ps_5}-\sqrt{2}(\ket{\ps_1}-\ket{\ps_2}+\ket{\ps_3}-\ket{\ps_4})).
 \eea
It is easy to verify that both $\r:=\sum^6_{i=1}\proj{\ps_i}$ 
and $\Gamma_1\r$ have rank 5 and that the 6 product states 
$\proj{\ps_i}$ are linearly independent. Since $\cR(\r)$ 
contains only 6 product vectors up to scalar multiples, 
$\r$ admits only one expression as a convex linear combination 
of normalized product states. Consequently, we have $L(\r)=6$.}
\eex

In connection with Proposition \ref{pp:kleq4}(i) above, we point 
out that $L(\r)=4$ does not imply that $\rank \G\r = 4$, 
$\forall \G\in\T$. A counterexample is the two-qubit 
separable state $I_4+(\ket{00}+\ket{11})(\bra{00}+\bra{11})$ 
of birank $(4,3)$ \cite{cd12pra}.

\section{Critical length}

Let us introduce three important integers associated to the length filtration.  
\bd
\label{df:ml}
First, the {\em maximum length} of separable states, 
$L_{\rm max}$, is the smallest positive integer $r$  such that $\cS'_r=\cS$. 
It follows from the definition of length that $\cS'_i$ is 
a proper subset of $\cS'_{i+1}$ for $i<L_{\max}$.

Second, the {\em critical length}, $L_{\rm crit}$, is the smallest positive integer $r$ such that $\dim \cS'_r=d^2-1$. 
Equivalently, it is the smallest $r$ such that $\cS'_r$ has 
positive volume as a subset of the affine hyperplane $H_1$.

Third, the integer $l$ introduced in our paper \cite{cd13jmp}. We rename it $L_c$, and recall its definition
\bea \label{eq:Lc-def}
L_c = \left\lceil \frac{d^2}{1+2\sum(d_i-1)} \right\rceil,
\eea
where $\left\lceil x \right\rceil$ denotes the least integer 
$k$ such that $x\le k$.
\ed

It was shown in \cite[Theorem 8]{cd13jmp} that there exist separable states of length $L_c$, and it follows from the same  theorem that 
\bea \label{eq:dim-duz}
\dim \cS'_r < d^2-1 \quad {\rm for} \quad r<L_c.
\eea

Let us show that $L_c\ge d$.
\bl \label{le:Lc-nejednakost}
We have $L_c\ge d$ and equality holds if and only if $n=2$ 
and $(d_1-2)(d_2-2)\le1$.
\el
\bpf
The inequality $L_c\ge d$ is equivalent to
$$
1+2\sum_{i=1}^n (d_i-1)<\frac{d^2}{d-1}=d+1+\frac{1}{d-1}.
$$
As $d>2$, this is equivalent to
\bea \label{eq:d-nejednakost}
d-2\sum_{i=1}^n (d_i-1) \ge 0.
\eea
In the proof of \cite[Corollary 9]{cd13jmp} it was shown that 
$f(d_1,\ldots,d_n):=d-2\sum_i (d_i-1)$ is a strictly increasing  function of a single variable $d_i$ (for each $i$). Hence, 
$f(d_1,\ldots,d_n)\ge f(2,\ldots,2)=2^n-2n\ge0$ and so  \eqref{eq:d-nejednakost} holds and we have $L_c\ge d$. 

Assume that $L_c=d$. Then we must have $2^n=2n$ and so $n=2$. By using  \eqref{eq:Lc-def} and $L_c\le d$, we obtain that 
$(d_1-2)(d_2-2)\le1$. Conversely, one can easily verify that $L_c=d$ when $n=2$ and $(d_1-2)(d_2-2)\le1$. 
\epf

To summarize, we have the following proposition.
\bpp 
\label{pp:lclcrit}
For any finite-dimensional multipartite quantum system, 
the following inequalities hold
\bea \label{eq:L-ineq}
d \le L_c \le L_{\rm crit} \le L_{\rm max} \le d^2.
\eea
\epp
The values of $L_{\rm max}$ are not known except for $d\le6$ in 
which case we have $L_{\rm max}=d$ ( see \cite{cd12pra}). 
In the systems $2\ox 4$ and $3\ox 3$ it is known that there exist separable states of length 10 \cite{hk13,hk14cmp}. 
By Lemma \ref{le:Lc-nejednakost}, in 
these two cases we have $L_c=d$ and so $L_c<10\le L_{\max}$. 
We single out the three smallest cases as an open and 
challenging problem.

{\bf Open problem 1}~ 
Find the value of $L_{\max}$ for the quantum systems 
$2\ox 4$, $3\ox 3$ and $2\ox 2\ox 2$.

Although the system $2\ox 4$ can be realized as the system 
$2\ox 2\ox 2$ of three qubits by grouping together 
the second and third parties, we do not know how the values 
of $L_{\max}$ in these two systems are related.
 
%

We have mentioned earlier that $\dim\cS=d^2-1$. However we believe that a much stronger claim holds.

\bcj \label{cj:Prva-hip}
For any finite-dimensional quantum system, we have
$\dim \cS'_{L_c}=d^2-1$ or, equivalently, $L_c=L_{\rm crit}$.
\ecj

We shall prove later (see Theorem \ref{thm:glavna}) that this conjecture is true in the bipartite case with one party being 
a single qubit (i.e., the case $n=d_1=2$ with arbitrary $d_2$). 
We have also verified the validity of this conjecture in the 
cases where the dimension vector $(d_1,d_2,\ldots,d_n)$ is one 
of the following: 
\begin{eqnarray*}
&& (2,2,2),~(2,2,2,2),~(2,2,2,2,2), \\
&& (2,2,3),~ (2,2,2,4),~ (2,2,2,5), \\
&& (3,3),~(3,4),~(3,5),~(3,6),~(3,7), \\
&& (4,4),~ (4,5),~ (4,6), \\
&& (5,5),~ (5,6).
\end{eqnarray*}
Some of these cases were handled in our paper 
\cite[Table 1]{cd13jmp}.

To simplify notation we set 
$\cH_\times=\cH_1\times\cH_2\times\cdots\times\cH_n$. 
Let $\ph:\cH_\times \to H$ be the map defined by
\bea
\label{eq:phz1}
\ph(z^{(1)},\ldots,z^{(n)})=
\proj{z^{(1)}}\ox\cdots\ox\proj{z^{(n)}}.
\eea

More generally, for any positive integer $r$ we shall define the map  $\Phi_r:\cH_\times^r \to H$. For convenience we write 
$z\in\cH_\times^r$ as an $r\times n$ matrix 
\bea
\label{eq:ziq}
z=[z^{(i,q)}]
\eea 
whose
rows are indexed by $i=1,2,\ldots,r$, the columns by
$q=1,2,\ldots,n$, and $\ket{z^{(i,q)}}\in\cH_q$ for each $i$ and $q$. We use the abbreviation $z^{(i)}$ for the $i$th row 
$(z^{(i,1)},\ldots,z^{(i,n)})$ of the matrix $z$.
Then $\Phi_r$ is defined by
\bea \label{eq:Phi_r}
\Phi_r(z) &=& \notag
\Phi_r(z^{(1)},\ldots,z^{(r)})=\sum_{i=1}^r \ph(z^{(i)}) \\
&=& \sum_{i=1}^r \proj{z^{(i,1)}}\ox\cdots\ox\proj{z^{(i,n)}}.
\eea
In the bipartite case these maps were introduced in 
\cite{cd13jmp}. 
It is obvious that $\Phi_r$ is invariant 
under permutations of the $z^{(i)}$.
We note that the range of $\Phi_r$ is the cone $K\cS'_r$. 
Since $\cS'_1$ is diffeomorphic to the product of the complex 
projective spaces ${\bf P}(\cH_i)\cong{\bf CP}^{d_i-1}$, we have 
$\dim K\cS'_1=1+2\sum(d_i-1)$. Hence, at the generic points 
$p\in\cH_\times$, we have
\bea \label{eq:rangJak}
\rank ({\rm d}\Phi_1)_p = 1+2\sum(d_i-1).
\eea

Let us illustrate the definition of $\Phi_r$ by a simple example.

\bex \label{ex:Jed-mat}
{\rm
In this example we consider the map $\Phi_d$, i.e., we set $r=d$. We choose a very special point $p\in\cH_\times^d$. In our matrix notation, $p$ is represented by the $d\times n$ matrix $[p^{(s,q)}]$, where $s$ runs through the set $S$ of all integral sequences $s=(s_1,\ldots,s_n)$ with $0\le s_i<d_i$, and 
$p^{(s,q)}=\ket{s_q}_q$. Note that $|S|=d$ and $\Phi_d(p)=I_d$, the identity operator. In particular, it follows that the state $I_d/d$ is separable and has length $d$.

When $n=2$ and $d_1=d_2=2$ (the two-qubit case) we have 
$S=\{(0,0),(0,1),(1,0),(1,1)\}$ and 
$$
p=\left[ \begin{array}{cc}
    \ket{0}_1 & \ket{0}_2\\
    \ket{0}_1 & \ket{1}_2\\
    \ket{1}_1 & \ket{0}_2\\
    \ket{1}_1 & \ket{1}_2 \end{array} \right].
$$
\qed
}
\eex

An intriguing question arises from the above example. As 
$d\le L_{\rm crit}$ and $I_d/d\in\cS'_d\sue\cS'_{L_{\rm crit}}$, 
it is plausible that $I_d/d$ belongs to the interior of 
$\cS'_{L_{\rm crit}}$.

\bcj \label{cj:Druga-hip}
For any finite-dimensional quantum system, the point $I_d/d$ 
lies in the interior of $\cS'_{L_{\rm crit}}$, i.e., 
there exists a small ball in $H_1$ centered at $I_d/d$ which is
contained in $\cS'_{L_{\rm crit}}$.
\ecj


\section{Rank of the Jacobian matrix of $\Phi_r$}
\label{sec:Jacobian}

Conjecture \ref{cj:Prva-hip} is equivalent to the assertion that 
the differential ${\rm d}\Phi_{L_c}$ generically has rank $d^2$. 
For that reason we shall compute the Jacobian matrix of $\Phi_r$ 
for any $r$.

We need to introduce the coordinates.
Let us write a vector $\ket{z^{(i,q)}}\in\cH_q$ as a linear combination of the basis vectors
\bea
\label{eq:ziq2}
\ket{z^{(i,q)}}=\sum_{j=0}^{d_q-1} \z^{(i,q)}_j \ket{j}_q.
\eea
By substituting these expressions into \eqref{eq:Phi_r}, we 
obtain 
\bea \notag
\Phi_r(z^{(1)},\ldots,z^{(r)}) &=&
\sum_{s=1}^r \ketbra{z^{(s,1)},\ldots,z^{(s,n)}}
{z^{(s,1)},\ldots,z^{(s,n)}} \\ 
\label{eq:vred-Fi}
&=& \sum_{{\bf j},{\bf k}} c({\bf j};{\bf k})
\ketbra{\bf j}{\bf k}, \\ \notag
{\bf j}:=j_1,\ldots,j_n; &~& {\bf k}:=k_1,\ldots,k_n;
\eea
where the indices $j_q$ and $k_q$ run from $0$ to $d_q-1$ for
each $q$, and the coefficients $c({\bf j},{\bf k})$ are given by
\bea \label{eq:coeff-c}
c({\bf j};{\bf k})=\sum_{s=1}^r \prod_{q=1}^n
\z^{(s,q)}_{j_q} \z^{(s,q)*}_{k_q}.
\eea
Note that $c({\bf j};{\bf k})$ is the $(j,k)$th entry of 
the $d\times d$ matrix \eqref{eq:vred-Fi}, where
\bea \notag
j=1+j_n+j_{n-1}d_n+j_{n-2}d_{n-1}d_n+\cdots+
j_1d_2d_3\cdots d_n, \\  \notag
k=1+k_n+k_{n-1}d_n+k_{n-2}d_{n-1}d_n+\cdots+
k_1d_2d_3\cdots d_n.
\eea

To introduce real coordinates, we shall write 
\bea \label{eq:real-coord}
\z^{(s,q)}_j=\x^{(s,q)}_j+{\bf i}\eta^{(s,q)}_j,
\eea
where $\x^{(s,q)}_j,\eta^{(s,q)}_j\in\bR$ and 
{\bf i} is the imaginary unit. 
One can easily verify that
\bea \label{eq:parc-izv-1}
\frac{ \partial }{ \partial\x^{(s,q)}_m }
\z^{(s,q)}_j \z^{(s,q)*}_k &=&
\delta_{mj}\z^{(s,q)*}_k + \delta_{mk} \z^{(s,q)}_j, 
\\ \label{eq:parc-izv-2}
\frac{ \partial }{ \partial\eta^{(s,q)}_m }
\z^{(s,q)}_j \z^{(s,q)*}_k &=& {\bf i} \left(
\delta_{mj}\z^{(s,q)*}_k - \delta_{mk} \z^{(s,q)}_j
\right).
\eea

Let $M'_r$ be the complex matrix whose rows are labeled with the $d^2$ symbols $({\bf j};{\bf k})$ and the columns with the 
$2r\sum d_q$ symbols $(s,q,j,\xi)$ and $(s,q,j,\eta)$, and the corresponding matrix entry is the partial derivative of 
$c({\bf j};{\bf k})$ with respect to the real variable 
$\x^{(s,q)}_j$ or $\eta^{(s,q)}_j$, respectively. 
So $M'_r$ is of size $d^2\times(2r\sum d_q)$. 
We denote by $[{\bf j};{\bf k}]'$ the row of $M'_r$ with 
label $({\bf j};{\bf k})$, and similarly let 
$[s,q,j,\xi]'$ and $[s,q,j,\eta]'$ denote the columns of $M'_r$ 
with labels $(s,q,j,\xi)$ and $(s,q,j,\eta)$, respectively.
We order the rows and the columns by using the lexicographic ordering of their labels, with the convention that $\xi<\eta$.

For instance, in the case $n=2$ with $d_1=d_2=2$ and $r=2$,
the 16 column labels are orderd as follows:
$(1,1,0,\xi)$, $(1,1,0,\eta)$, 
$(1,1,1,\xi)$,  $(1,1,1,\eta)$, 
$(1,2,0,\xi)$, $(1,2,0,\eta)$, 
$(1,2,1,\xi)$,  $(1,2,1,\eta)$, 
$(2,1,0,\xi)$, $(2,1,0,\eta)$, 
$(2,1,1,\xi)$,  $(2,1,1,\eta)$, 
$(2,2,0,\xi)$, $(2,2,0,\eta)$, 
$(2,2,1,\xi)$,  $(2,2,1,\eta)$.

Since the matrix \eqref{eq:vred-Fi} is Hermitian, the rows 
$[{\bf j};{\bf k}]'$ and $[{\bf k};{\bf j}]'$ are complex 
conjugates of each other. 

Let $M_r$ be the matrix obtained from 
$M'_r$ by the following substitutions: if (lexicographically) 
${\bf j}<{\bf k}$ resp. ${\bf j}>{\bf k}$  
then we replace each entry in the row $[{\bf j};{\bf k}]'$ with its real resp. imaginary  part. Then $M_r$ is the Jacobian matrix 
of ${\rm d}\Phi_r$ (with respect to a suitable basis of $H$). 
The rows and columns of $M_r$ will be denoted in the same way 
as for $M'_r$ except that we will omit the apostrophe sign.

It follows from \eqref{eq:coeff-c} and \eqref{eq:parc-izv-1} that the entry of $M'_r$ in row $[{\bf j};{\bf k}]'$ and column 
$[s,t,m,\xi]'$ is equal to
\bea
\label{eq:deltamjt}
\left(
\delta_{m,j_t} \z^{(s,t)*}_{k_t} + 
\delta_{m,k_t} \z^{(s,t)}_{j_t}
\right)
\prod_{q\ne t} \z^{(s,q)}_{j_q} \z^{(s,q)*}_{k_q}.
\eea
Similarly, the entry in row $({\bf j};{\bf k})$ and column 
$[s,t,m,\eta]'$ is equal to
\bea
\label{eq:ideltamjt}
{\bf i} \left(
\delta_{m,j_t} \z^{(s,t)*}_{k_t} - 
\delta_{m,k_t} \z^{(s,t)}_{j_t}
\right)
\prod_{q\ne t} \z^{(s,q)}_{j_q} \z^{(s,q)*}_{k_q}.
\eea

In the special case $r=1$ the matrix $M'_1$ depends only 
on the variables $\z^{(1,q)}_j$, where $q=1,\ldots,n$ and
$j=0,1,\ldots d_q -1$. So $M'_1$ has $d^2$ rows and $2\sum d_q$ 
columns. We indicate this dependence by 
writing $M'_1$ as $M'_1( \z^{(1,q)}_j )$. Then $M'_r$ has a 
very simple expression, namely
\bea \label{eq:Matrica-M}
M'_r=\left[ M'_1( \z^{(1,q)}_j ) ~  M'_1( \z^{(2,q)}_j ) ~ 
\cdots ~ M'_1( \z^{(r,q)}_j ) \right].
\eea
This reduces the problem of computing $M'_r$ to the computation 
of $M'_1$ (and the same is valid for $M_r$ and $M_1$).

Thus we have explicit formulas for the entries of the matrices 
$M'_r$ and $M_r$ for any $r$. Let us give an explicit example.

\bex \label{ex:2qPhi_1}
{\rm
In the case of two qubits the matrix $M'_1$ has size $16\times8$. As $r=1$, we must have $s=1$. Thus, in displaying this matrix below we may omit the first superscript:


\begin{footnotesize}
\begin{equation*}
\left[
\begin{array}{cccccccc}
2\x^{(1)}_0 |\z^{(2)}_0|^2 & 
2\eta^{(1)}_0 |\z^{(2)}_0|^2 & 
0 & 
0 & 
2|\z^{(1)}_0 |^2 \x^{(2)}_0 & 
2|\z^{(1)}_0|^2 \eta^{(2)}_0 & 
0 &
0 \\                                           
2\x^{(1)}_0 \z^{(2)}_0 \z^{(2)*}_1 & 
2\eta^{(1)}_0 \z^{(2)}_0 \z^{(2)*}_1 & 
0 & 
0 & 
|\z^{(1)}_0|^2 \z^{(2)*}_1 & 
{\bf i} |\z^{(1)}_0|^2 \z^{(2)*}_1 & 
|\z^{(1)}_0|^2 \z^{(2)}_0 & 
-{\bf i} |\z^{(1)}_0|^2 \z^{(2)}_0 \\           
\z^{(1)*}_1 |\z^{(2)}_0|^2 & 
{\bf i} \z^{(1)*}_1 |\z^{(2)}_0|^2 &
\z^{(1)}_0 |\z^{(2)}_0|^2 & 
-{\bf i} \z^{(1)}_0 |\z^{(2)}_0|^2 & 
2\z^{(1)}_0 \z^{(1)*}_1 \x_0^{(2)} & 
2\z^{(1)}_0 \z^{(1)*}_1 \eta^{(2)}_0 & 
0 & 
0 \\                                             
\z^{(1)*}_1 \z^{(2)}_0 \z^{(2)*}_1 &
{\bf i} \z^{(1)*}_1 \z^{(2)}_0 \z^{(2)*}_1 &
\z^{(1)}_0 \z^{(2)}_0 \z^{(2)*}_1 &
-{\bf i} \z^{(1)}_0 \z^{(2)}_0 \z^{(2)*}_1 &
\z^{(1)}_0 \z^{(1)*}_1 \z^{(2)*}_1 &
{\bf i} \z^{(1)}_0 \z^{(1)*}_1 \z^{(2)*}_1 &
\z^{(1)}_0 \z^{(1)*}_1 \z^{(2)}_0 &
-{\bf i} \z^{(1)}_0 \z^{(1)*}_1 \z^{(2)}_0 \\    
2\x^{(1)}_0 \z^{(2)*}_0 \z^{(2)}_1 & 
2\eta^{(1)}_0 \z^{(2)*}_0 \z^{(2)}_1 & 
0 & 
0 & 
|\z^{(1)}_0|^2 \z^{(2)}_1 & 
-{\bf i} |\z^{(1)}_0|^2 \z^{(2)}_1 & 
|\z^{(1)}_0|^2 \z^{(2)*}_0 &
{\bf i} |\z^{(1)}_0|^2 \z^{(2)*}_0 \\             
2\x^{(1)}_0 |\z^{(2)}_1|^2 & 
2\eta^{(1)}_0 |\z^{(2)}_1|^2 & 
0 & 
0 & 
0 & 
0 & 
2|\z^{(1)}_0|^2 \x^{(2)}_1 &
2|\z^{(1)}_0|^2 \eta^{(2)}_1 \\                   
\z^{(1)*}_1 \z^{(2)*}_0 \z^{(2)}_1 &
{\bf i} \z^{(1)*}_1 \z^{(2)*}_0 \z^{(2)}_1 &
\z^{(1)}_0 \z^{(2)*}_0 \z^{(2)}_1 &
-{\bf i} \z^{(1)}_0 \z^{(2)*}_0 \z^{(2)}_1 &
\z^{(1)}_0 \z^{(1)*}_1 \z^{(2)}_1 &
-{\bf i} \z^{(1)}_0 \z^{(1)*}_1 \z^{(2)}_1 &
\z^{(1)}_0 \z^{(1)*}_1 \z^{(2)*}_0 &
{\bf i} \z^{(1)}_0 \z^{(1)*}_1 \z^{(2)*}_0 \\      
\z^{(1)*}_1 |\z^{(2)}_1|^2 &
{\bf i} \z^{(1)*}_1 |\z^{(2)}_1|^2 & 
\z^{(1)}_0 |\z^{(2)}_1|^2 & 
-{\bf i} \z^{(1)}_0 |\z^{(2)}_1|^2 & 
0 & 
0 & 
2\z^{(1)}_0 \z^{(1)*}_1 \x^{(2)}_1 & 
2\z^{(1)}_0 \z^{(1)*}_1 \eta^{(2)}_1 \\            
\z^{(1)}_1 |\z^{(2)}_0|^2 & 
-{\bf i} \z^{(1)}_1 |\z^{(2)}_0|^2 & 
\z^{(1)*}_0 |\z^{(2)}_0|^2 &
{\bf i} \z^{(1)*}_0 |\z^{(2)}_0|^2 & 
2\z^{(1)*}_0 \z^{(1)}_1 \x^{(2)}_0 & 
2\z^{(1)*}_0 \z^{(1)}_1 \eta^{(2)}_0 & 
0 &
0 \\                                               
\z^{(1)}_1 \z^{(2)}_0 \z^{(2)*}_1 &
-{\bf i} \z^{(1)}_1 \z^{(2)}_0 \z^{(2)*}_1 &
\z^{(1)*}_0 \z^{(2)}_0 \z^{(2)*}_1 &
{\bf i} \z^{(1)*}_0 \z^{(2)}_0 \z^{(2)*}_1 &
\z^{(1)*}_0 \z^{(1)}_1 \z^{(2)*}_1 &
{\bf i} \z^{(1)*}_0 \z^{(1)}_1 \z^{(2)*}_1 &
\z^{(1)*}_0 \z^{(1)}_1 \z^{(2)}_0 &
-{\bf i} \z^{(1)*}_0 \z^{(1)}_1 \z^{(2)}_0 \\      
0 & 
0 & 
2\x^{(1)}_1 |\z^{(2)}_0|^2 &
2\eta^{(1)}_1 |\z^{(2)}_0|^2 & 
2|\z^{(1)}_1|^2 \x^{(2)}_0 & 
2|\z^{(1)}_1|^2 \eta^{(2)}_0 & 
0 & 
0 \\                                               
0 & 
0 & 
2\x^{(1)}_1 \z^{(2)}_0 \z^{(2)*}_1 & 
2\eta^{(1)}_1 \z^{(2)}_0 \z^{(2)*}_1 &
|\z^{(1)}_1|^2 \z^{(2)*}_1 & 
{\bf i} |\z^{(1)}_1|^2 \z^{(2)*}_1 & 
|\z^{(1)}_1|^2 \z^{(2)}_0 & 
-{\bf i} |\z^{(1)}_1|^2 \z^{(2)}_0 \\              
\z^{(1)}_1 \z^{(2)*}_0 \z^{(2)}_1 &
-{\bf i} \z^{(1)}_1 \z^{(2)*}_0 \z^{(2)}_1 &
\z^{(1)*}_0 \z^{(2)*}_0 \z^{(2)}_1 &
{\bf i} \z^{(1)*}_0 \z^{(2)*}_0 \z^{(2)}_1 &
\z^{(1)*}_0 \z^{(1)}_1 \z^{(2)}_1 &
-{\bf i} \z^{(1)*}_0 \z^{(1)}_1 \z^{(2)}_1 &
\z^{(1)*}_0 \z^{(1)}_1 \z^{(2)*}_0 &
{\bf i} \z^{(1)*}_0 \z^{(1)}_1 \z^{(2)*}_0 \\       
\z^{(1)}_1 |\z^{(2)}_1|^2 &
-{\bf i} \z^{(1)}_1 |\z^{(2)}_1|^2 &
\z^{(1)*}_0 |\z^{(2)}_1|^2 & 
{\bf i} \z^{(1)*}_0 |\z^{(2)}_1|^2 & 
0 & 
0 & 
2\z^{(1)*}_0 \z^{(1)}_1 \x^{(2)}_1 & 
2\z^{(1)*}_0 \z^{(1)}_1 \eta^{(2)}_1 \\             
0 & 
0 & 
2\x^{(1)}_1 \z^{(2)*}_0 \z^{(2)}_1 & 
2\eta^{(1)}_1 \z^{(2)*}_0 \z^{(2)}_1 & 
|\z^{(1)}_1|^2 \z^{(2)}_1 & 
-{\bf i} |\z^{(1)}_1|^2 \z^{(2)}_1 & 
|\z^{(1)}_1|^2 \z^{(2)*}_0 &
{\bf i} |\z^{(1)}_1|^2 \z^{(2)*}_0 \\               

0 & 
0 & 
2\x^{(1)}_1 |\z^{(2)}_1|^2 & 
2\eta^{(1)}_1 |\z^{(2)}_1|^2 & 
0 & 
0 & 
2|\z^{(1)}_1 |^2 \x^{(2)}_1 & 
2|\z^{(1)}_1|^2 \eta^{(2)}_1 \\                     
\end{array}
\right].
\end{equation*}
\end{footnotesize}
Let us evaluate the matrices $M'_1$ and $M_1$ at the point 
$p=[ \ket{0}_1 ~ \ket{0}_2]$. Except for 
$\x^{(1)}_0=\x^{(2)}_0=1$, all other coordinates of $p$ vanish.
By dropping the zero rows, we obtain the matrices
$$
\begin{array}{c}
00;00\\ 00;01\\ 00;10\\ 01;00\\ 10;00
\end{array}\quad
\left[ \begin{array}{cccccccc}
2&0&0&0&2&0&0&0\\
0&0&0&0&0&0&1&-{\bf i}\\
0&0&1&-{\bf i}&0&0&0&0\\
0&0&0&0&0&0&1&{\bf i}\\
0&0&1&{\bf i}&0&0&0&0
\end{array} \right],\quad
\left[ \begin{array}{cccccccc}
2&0&0&0&2&0&0&0\\
0&0&0&0&0&0&1&0\\
0&0&1&0&0&0&0&0\\
0&0&0&0&0&0&0&1\\
0&0&0&1&0&0&0&0
\end{array} \right].
$$
On the left of the matrices we show the row labels inherited 
from $M'_1$.
It is obvious that $M_1$ has rank 5, i.e., the rank 
of ${\rm d}\Phi_1$ at $p$ is 5. This agrees with the 
general formula \eqref{eq:rangJak}.
\qed
}
\eex

\section{The bipartite case $2\ox N$} \label{sec:2xN}

In this section we specialize to the bipartite case 
$2\otimes N$. 
Thus we set $n=2$, $d_1=2$, $d_2=N$, and so $d=2N$. 
Further, we set $r=2N$ and write $M'$ and $M$ instead of 
$M'_r$ and $M_r$, respectively. 
Our objective is to prove that Conjecture \ref{cj:Prva-hip} 
is true in this case. In fact we shall prove that generically $M_r$ has rank $d=4N^2$.

Let $a_i,b_i$, $i=1,\ldots,N$, be real parameters and $p$ the 
point in $\cH_\times^{2N}$ given by the matrix
\bea
\label{eq:a}
\left[ \begin{array}{ll}
\ket{0}_1 +        a_1\ket{1}_1  & \ket{0}_2 \\
\ket{0}_1 + {\bf i}b_1\ket{1}_1  & \ket{0}_2 \\
\ket{0}_1 +        a_2\ket{1}_1  & \ket{1}_2 \\
\ket{0}_1 + {\bf i}b_2\ket{1}_1  & \ket{1}_2 \\
\ket{0}_1 +        a_3\ket{1}_1  & \ket{2}_2 \\
\ket{0}_1 + {\bf i}b_3\ket{1}_1  & \ket{2}_2 \\
\vdots                        & \vdots\\
\ket{0}_1 +        a_N\ket{1}_1  & \ket{N-1}_2 \\
\ket{0}_1 + {\bf i}b_N\ket{1}_1  & \ket{N-1}_2 
\end{array} \right].
\eea
Thus the $\z$-coordinates of $p$ are $\z^{(s,1)}_0=1$ for  
$s=1,2,\ldots,2N$; $\z^{(2i-1,1)}_1=a_i$ and 
$\z^{(2i,1)}_1={\bf i}b_i$ for $i=1,2,\ldots,N$; 
$\z^{(2i-1,2)}_j=\z^{(2i,2)}_j=\delta_{j,i-1}$ 
for $i=1,2,\ldots,N$ and  $j=0,1,\ldots,N-1$.

We shall evaluate the matrix $M'$ at the point $p$.  It 
has $d^2$ rows and $4N(2+N)$ columns. 
The row labels are $(j_1,j_2;k_1,k_2)$ where 
$j_1,k_1\in\{0,1\}$ and $j_2,k_2\in\{0,1,\ldots,N-1\}$.
The column labels are $(s,t,m,\xi)$ and $(s,t,m,\eta)$ 
where $s\in\{1,2,\ldots,2N\}$, $t\in\{1,2\}$, and 
$m\in\{0,1\}$ if $t=1$ while $m\in\{0,1,\ldots,N-1\}$ if $t=2$. 
For a given $s$, we define $s'$ by writing $s=2s'-1$ 
if $s$ is odd and $s=2s'$ if $s$ is even.
 
For each column of $M'$ and each nonzero entry in that 
column, we list first the row label $(j_1,j_2;k_1,k_2)$ where 
this entry occurs and then the entry itself. 
All non-listed entries are 0.
The entries are computed by using the formulas 
\eqref{eq:deltamjt} and \eqref{eq:ideltamjt}.

Case 1: $[s,1,m,\xi]'$.
$$
m=0: \quad
\begin{array}{ccc}
{\rm row} & s~ {\rm odd} & s~ {\rm even} \\
\hline
(0,s'-1;0,s'-1) & 2 & 2 \\
(0,s'-1;1,s'-1) & a_{s'} & -{\bf i} b_{s'} \\
(1,s'-1;0,s'-1) & a_{s'} & {\bf i} b_{s'} 
\end{array}
$$
$$
m=1: \quad
\begin{array}{ccc}
{\rm row} & s~ {\rm odd} & s~ {\rm even} \\
\hline
(0,s'-1;1,s'-1) & 1 & 1 \\
(1,s'-1;0,s'-1) & 1 & 1 \\
(1,s'-1;1,s'-1) & 2a_{s'} & 0 
\end{array}
$$

Case 2: $[s,1,m,\eta]'$.
$$
m=0: \quad
\begin{array}{ccc}
{\rm row} & s~ {\rm odd} & s~ {\rm even} \\
\hline
(0,s'-1;1,s'-1) & {\bf i} a_{s'} &  b_{s'} \\
(1,s'-1;0,s'-1) & -{\bf i} a_{s'} &  b_{s'} 
\end{array}
$$
$$
m=1: \quad
\begin{array}{ccc}
{\rm row} & s~ {\rm odd} & s~ {\rm even} \\
\hline
(0,s'-1;1,s'-1) & -{\bf i} & -{\bf i} \\
(1,s'-1;0,s'-1) & {\bf i} & {\bf i} \\
(1,s'-1;1,s'-1) & 0  & 2b_{s'}
\end{array}
$$

Case 3: $[s,2,m,\xi]'$.
$$
m=s'-1: \quad
\begin{array}{ccc}
{\rm row} & s~ {\rm odd} & s~ {\rm even} \\
\hline
(0,s'-1;0,s'-1) & 2 & 2 \\
(0,s'-1;1,s'-1) & 2a_{s'} & -2{\bf i} b_{s'} \\
(1,s'-1;0,s'-1) & 2a_{s'} & 2{\bf i} b_{s'} \\
(1,s'-1;1,s'-1) & 2a_{s'}^2 & 2b_{s'}^2 
\end{array}
$$
In the next table $\{j_2,k_2\}=\{m,s'-1\}$.
$$
m\ne s'-1: \quad
\begin{array}{ccc}
{\rm row} & s~ {\rm odd} & s~ {\rm even} \\
\hline
(0,j_2;0,k_2) & 1 & 1 \\
(0,j_2;1,k_2) & a_{s'} & -{\bf i} b_{s'} \\
(1,j_2;0,k_2) & a_{s'} & {\bf i} b_{s'} \\
(1,j_2;1,k_2) & a_{s'}^2 & b_{s'}^2
\end{array}
$$

Case 4: $[s,2,m,\eta]'$.
$$
m\ne s'-1: \quad
\begin{array}{ccc}
{\rm row} & s~ {\rm odd} & s~ {\rm even} \\
\hline
(0,m;0,s'-1) & {\bf i} & {\bf i} \\
(0,m;1,s'-1) & {\bf i} a_{s'} &  b_{s'} \\
(1,m;0,s'-1) & {\bf i} a_{s'} & -b_{s'} \\
(1,m;1,s'-1) & {\bf i} a_{s'}^2 & {\bf i} b_{s'}^2 \\
(0,s'-1;0,m) & -{\bf i} & -{\bf i} \\
(0,s'-1;1,m) & -{\bf i} a_{s'} & -b_{s'} \\
(1,s'-1;0,m) & -{\bf i} a_{s'} &  b_{s'} \\
(1,s'-1;1,m) & -{\bf i} a_{s'}^2 & -{\bf i} b_{s'}^2
\end{array}
$$

Note that in the case 4) only $m\ne s'-1$ is shown. This 
means that the $2N$ columns $[s,2,s'-1,\eta]'$ of $M'$ are 0. 
Consequently, the $2N$ columns $[s,2,s'-1,\eta]$ of $M$ are 
also 0. Let $M^\#$ be the square matrix of order $4N^2$ which 
is obtained from $M$ by removing these $2N$ zero columns 
and the additional $6N$ columns with labels $[s,1,0,\eta]$, 
$[s,2,s'-1,\xi]$ for $s=1,2,\ldots,2N$ and $[s,1,0,\xi]$ 
and $[s,1,1,\eta]$ for $s=2,4,\ldots,2N$. 

We can now prove our main result which shows that 
Conjecture \ref{cj:Prva-hip} is valid in $2\ox N$.

\bt \label{thm:glavna} In the bipartite system $2\ox N$, 
we have $2N=L_c=L_{\rm crit}$. Equivalently, 
$\dim \cS'_{2N}=4N^2-1$.
\et

\bpf
It suffices to show that generically the matrix $M$ has 
rank $4N^2$. We shall prove the stronger assertion, namely that 
\bea \label{eq:detMr}
\det M^\# &=& \pm 2^{N(N+1)} \prod_{q=1}^N a_q\cdot
\left( \prod_{i<j}(a_i-a_j)(b_i-b_j)(a_ia_j-b_ib_j) \right)^2.
\eea 

To avoid confusion, we shall refer to the rows and the columns of $M^\#$ by the labels inherited from $M$. 

The columns $[s,1,1,\eta]$ for $s$ odd and $[s,1,1,\xi]$ for 
$s$ even belong to $M^\#$ and have exactly one nonzero entry. 
This entry is equal to 1 and occurs in the row 
$(1,s'-1;0,s'-1)$ and $(0,s'-1;1,s'-1)$, respectively. 
Let us remove from $M^\#$ these $2N$ rows and $2N$ columns. 
Then, in the remaining matrix, each of the rows with the 
diagonal labels, i.e., labels having the form 
$(j_1,j_2;j_1,j_2)$, has a single nonzero entry. 
This entry is in the column $[2j_2+1,1,j_1,\xi]$ and is 
equal to 2 if $j_1=0$ and to $2a_{j_2}$ if $j_1=1$. 
Let $M^{\#\#}$ be the matrix of order $4N(N-1)$ obtained 
by removing from $M^\#$ also these additional $2N$ rows and 
$2N$ columns. It follows that
\bea \label{eq:detMrr}
\det M^\#=\pm 2^{2N} \prod_{q=1}^N a_q \cdot \det M^{\#\#}.
\eea 

One can verify easily that the rows of $M^{\#\#}$ have 
the labels $(j_1,j_2;k_1,k_2)$ where $j_2\ne k_2$, and 
that the columns of $M^{\#\#}$ have the labels $(s,2,m,w)$ 
where $s\in\{1,2,\ldots,2N\}$, $m\in\{0,1,\ldots,N-1\}$ 
with $m\ne s'-1$, and $w\in\{\xi,\eta\}$.

Let $u,v$ be integers such that $0\le u<v<N$. We define 
$R_{u,v}$ to be the set of 8 row labels $(j_1,u;k_1,v)$ and 
$(j_1,v;k_1,u)$. We define $C_{u,v}$ to be the set of 8 column 
labels $(2u+1,2,v,w)$, $(2u+2,2,v,w)$, $(2v+1,2,u,w)$, $(2v+2,2,u,w)$ where $w\in\{\xi,\eta\}$. The sets $R_{u,v}$ form 
a partition of the set of row labels of $M^{\#\#}$. 
Similarly, the sets $C_{u,v}$ form a partition of the set of 
column labels of $M^{\#\#}$. 
We denote by $M^{\#\#}_{u,v}$ the $8\times8$ submatrix of 
$M^{\#\#}$ with row labels $R_{u,v}$ and column labels 
$C_{u,v}$.

If $(s,2,m,w)\in C_{u,v}$ we claim that all nonzero 
entries of that column of $M^{\#\#}$ lie in the rows with 
label in $R_{u,v}$. Let us verify this claim for the column 
$(s=2u+1,2,m=v,\xi)\in C_{u,v}$ of $M^{\#\#}$. In this case 
we have $s'-1=u<v=m$. The nonzero entries in $M'$ lying 
in the column with label $(s,2,m,\xi)$ are listed in the 
second table of Case 3 above. As $s$ is odd, these entries
belong to $\{1,a_{u+1},a_{u+1}^2\}$. In particular, they are 
real. By using this table, we find that the nonzero 
entries of $M$ in column $(2u+1,2,v,\xi)$ are 
1 in row $(0,u;0,v)$, $a_{u+1}$ in rows $(0,u;1,v)$ and 
$(0,v;1,u)$, and $a_{u+1}^2$ in row $(1,u;1,v)$. Observe that 
these four rows indeed belong to $R_{u,v}$. 
Similarly, we find that the nonzero entries of column 
$(2v+1,2,u,\xi)$ are 1 in row $(0,u;0,v)$, $a_{v+1}$ in rows 
$(0,u;1,v)$ and $(0,v;1,u)$, and $a_{u+1}^2$ in row $(1,u;1,v)$.
We omit this verification for the other six columns in 
$C_{u,v}$. 

From the above claim it follows that, up to row and column 
permutations, $M^{\#\#}$ is the direct sum of the 
$N(N-1)/2$ blocks $M^{\#\#}_{u,v}$. Consequently, we have
\bea \label{eq:detMuv}
\det M^{\#\#}=\pm \prod_{0\le u<v<N} \det M^{\#\#}_{u,v}. 
\eea

Next one can show that, up to row and column permutations, 
each block $M^{\#\#}_{u,v}$ has the following simple form:
\begin{equation} \label{eq:8x8block}
\begin{array}{cc}
0,u;0,v & \quad \\
0,u;1,v & \quad \\
0,v;1,u & \quad \\
1,u;1,v & \quad \\
0,v;0,u & \\
1,u;0,v & \\
1,v;0,u & \\
1,v;1,u &
\end{array}
\left[ \begin{array}{cccccccc}
1 & 1 & 0 & 0 & 1 & 1 & 0 & 0  \\
a & a'& 0 & 0 & 0 & 0 & -b & -b' \\
a & a'& 0 & 0 & 0 & 0 & b & b' \\
a^2 & {a'}^2 & 0 & 0 & b^2 & {b'}^2 & 0 & 0 \\
0 & 0 & 1 & -1 & 0 & 0 & 1 & 1 \\
0 & 0 & -a & a' & b & b' & 0 & 0 \\
0 & 0 & a & -a' & b & b' & 0 & 0 \\
0 & 0 & a^2 & -{a'}^2 & 0 & 0 & b^2 & {b'}^2 
\end{array} \right],
\end{equation}
where $a=a_{u+1}$, $a'=a_{v+1}$, $b=b_{u+1}$ and $b'=b_{v+1}$. 
On the left of this matrix we show the labels of the rows.
The first two columns of the matrix \ref{eq:8x8block} have been 
computed above. The labels of the columns 1-8 of this matrix are 
$(2u+1,2,v,\xi)$, 
$(2v+1,2,u,\xi)$,
$(2u+1,2,v,\eta)$,
$(2v+1,2,u,\eta)$,
$(2u+2,2,v,\xi)$, 
$(2v+2,2,u,\xi)$, 
$(2u+2,2,v,\eta)$,
$(2v+2,2,u,\eta)$ respectively.
Consequently,
$$
\det  M^{\#\#}_{u,v} = \pm 4(a_{u+1}-a_{v+1})^2
(b_{u+1}-b_{v+1})^2 (a_{u+1}a_{v+1}-b_{u+1}b_{v+1})^2.
$$
Now the formula \eqref{eq:detMr} follows from 
\eqref{eq:detMrr} and \eqref{eq:detMuv}.
\epf



\section*{Acknowledgments}

LC was supported by the NSF of China (Grant No. 11501024), and the Fundamental Research Funds for the Central Universities (Grant Nos. 30426401 and 30458601).  
The second author was supported in part by an NSERC Discovery 
Grant.


\end{document}